*Dedicated to Professor Valentin I. Vlad's 70<sup>th</sup> Anniversary*

# CONTROL OF HIGH POWER PULSES EXTRACTED FROM THE MAXIMALLY COMPRESSED PULSE IN A NONLINEAR OPTICAL FIBER

GUANGYE YANG,[1,2] LU LI,[1,*] SUOTANG JIA,[3] DUMITRU MIHALACHE[4,5]

[1] Institute of Theoretical Physics, Shanxi University, Taiyuan, 030006, China
[2] Department of Physics, Shanxi Medical University, Taiyuan, 030001, China
[3] College of Physics and Electronics Engineering, Shanxi University, Taiyuan, 030006, China
[4] Academy of Romanian Scientists, 54 Splaiul Independentei, RO-050094 Bucharest, Romania
[5] "Horia Hulubei" National Institute of Physics and Nuclear Engineering, Magurele-Bucharest, Romania
*E-mail: llz@sxu.edu.cn

*Abstract.* We address the possibility to control high power pulses extracted from the maximally compressed pulse in a nonlinear optical fiber by adjusting the initial excitation parameters. The numerical results show that the power, location and splitting order number of the maximally compressed pulse and the transmission features of high power pulses extracted from the maximally compressed pulse can be manipulated through adjusting the modulation amplitude, width, and phase of the initial Gaussian-type perturbation pulse on a continuous wave background.

*Key words*: high power pulses, maximally compressed pulse, nonlinear Schrödinger equation.

## 1. INTRODUCTION

The nonlinear Schrödinger (NLS) equation is a central model describing a variety of nonlinear localization effects and has been extensively applied in various areas of nonlinear science, such as hydrodynamics, plasma physics, optics and photonics and so on [1–3]. Based on this model, various types of soliton solutions on a finite background, such as the Kuznetsov-Ma soliton, the Akhmediev breather and the Peregrine solution, have been obtained [4–9]. Among these solutions, the Kuznetsov-Ma soliton exhibits a periodization process of the bright soliton and the Akhmediev breather features a localized process of the continuous wave (CW) [10, 11], while the Peregrine solution as the limitation of the Kuznetsov-Ma soliton and the Akhmediev breather presents a localized wave-packet in both time and space, which is recognized as a prototype of *rogue wave* events (it is also called the Peregrine rogue wave) and can be used to generate highly energetic optical pulses from supercontinuum generation and optical turbulence [12–17]; see Ref. [18] for an overview of recent progress in the area of optical rogue waves in different physical settings.

Recently, the rogue wave phenomenon in various physical models has been intensively studied, including nonautonomous "rogons" in the inhomogeneous NLS equation with variable coefficients [19], $N$-order bright and dark rogue waves in resonant erbium-doped fiber systems [20], rogue-wave solutions of a three-component coupled NLS equation [21], rogue waves of the Hirota and the Maxwell-Bloch equations [22], the persistence of rogue waves in the integrable Sasa-Satsuma equation [23–26], the general rogue waves in the Davey-Stewartson I equation [27], and the study of rogue wave solutions and interactions for the generalized higher-order NLS equation with space- and time-modulated parameters [28]. Other recent relevant works in this broad area deal with the study of modulation instabilities of CW backgrounds and the formation of both solitons and rogue waves in a parity-time (PT)-symmetric system of linearly coupled nonlinear waveguides [29], dissipative rogue wave generation in mode-locked fiber lasers [30], electromagnetic rogue waves in beam–plasma interactions [31], and the study of the so-called *optical tsunamis* in nonlinear optical fibers with normal dispersion [32]. Such studies might be extended to other relevant dynamical models describing the formation of ultrashort temporal solitons in the femtosecond domain; see a few recent overviews of several non-NLS-type models for the description of ultrashort optical solitons which form and are quite robust in a variety of physical settings [33]. Relevant experiments on the formation mechanism and dynamics of the Peregrine rogue wave in optical fibers, in water wave tanks and in plasma have been recently reported [34–38].

It should be pointed out that the Akhmediev breather is a family of exact periodic solutions in transverse dimension for NLS equation [8], which can describe the evolution of initially continuous waves and can be excited by a weak periodic modulation [39]. Based on the Akhmediev breather, the continuous wave supercontinuum generation from modulation instability and Fermi-Pasta-Ulam (FPU) recurrence has been examined [40–42]. In contrast to the Akhmediev breather, the Kuznetsov-Ma soliton is a localized solution in transverse dimension for NLS equation, and can be excited by strongly modulated continuous wave [36] or by a small localized (single peak) perturbation pulse of CW background [11].

As possible applications, we recently advanced the idea that the Peregrine rogue wave can be used to generate high power pulses at the maximally compressed position in optical fiber [43]. Motivated by that idea, in this paper, we will study the evolution dynamics of the maximally compressed pulse induced by the Peregrine rogue wave over a wider range of initial conditions in optical fibers, and we show how to control the generation of such high power pulses.

The paper is organized as follows. In Sec. 2 we discuss in detail the process of excitation of the maximally compressed pulse from an input consisting of a superposition of a CW background and a Gaussian-type perturbation pulse. We show that it is possible to control the power, location and splitting order number of the maximally compressed pulse. The transmission characteristics of high power pulses extracted from the maximally compressed pulse can be adequately manipulated through adjusting the modulation amplitude, width, and phase of the input pulse. Our results are briefly summarized in Sec. 3.

## 2. EXCITATION OF MAXIMALLY COMPRESSED PULSE AND CONTROL OF GENERATION OF HIGH POWER PULSES

The propagation of optical pulses inside single-mode nonlinear fibers can be theoretically described by the NLS equation as follows [3]

$$\mathrm{i}\frac{\partial Q}{\partial \xi} - \frac{\beta_2}{2}\frac{\partial^2 Q}{\partial \tau^2} + \gamma |Q|^2 Q = 0, \tag{1}$$

where $Q(\xi,\tau)$ is the slowly varying amplitude of the pulse envelope, $\xi$ is the propagation distance and $\tau$ is the retarded time in a frame of reference moving with the pulse at the group velocity $v_g$ ($\tau = t - \xi/v_g$).

Equation (1) has a rational fraction solution that is called the Peregrine solution or the Peregrine rogue wave, which is a superposition of a CW solution of Eq. (1) and a rational fraction function and forms a maximally compressed pulse at the peak position [4,43]. This implies that the Peregrine rogue wave or the maximally compressed pulse can be excited by a small localized (single peak) perturbation pulse of CW background, see [43] for more details. To further study the initial excitation of the maximally compressed pulse, we here consider a more practical initial condition as follows

$$Q(0,\tau) = \sqrt{P_0}\left[1 + \delta \exp(-\tau^2/2T_1^2)\mathrm{e}^{\mathrm{i}\varphi}\right], \tag{2}$$

that is a superposition of a CW background and a Gaussian-type perturbation pulse, where $P_0$ is the initial background power, $\delta$ is a modulation amplitude, $T_1$ and $\varphi$ are the width and phase of the Gaussian-type perturbation pulse, respectively.

The extensive numerical simulations have shown that when such initial modulated field propagates along the fiber, it undergoes a dynamical nonlinear compression to yield a maximally compressed pulse at the peak position, which corresponds to the occurrence of the maximal peak power, and then is split into sub-pulses, see Ref. [43]. In the following, we investigate the influence of these initial parameters on the dynamics of the maximally compressed pulse.

First, we consider the influence of the modulation amplitude $\delta$ and of the width $T_1$ of the initial Gaussian-type perturbation pulse on the dynamics of the maximally compressed pulse for a given initial phase $\varphi = 0$. Here we used realistic parameters of a commercially available photonic crystal fiber (standard silica SMF-28 fiber) at the wavelength 1550 nm with the group velocity dispersion $\beta_2 = -21.4$ ps²/km, and the nonlinear parameter $\gamma = 1.2$ W$^{-1}\cdot$km$^{-1}$ [35, 44].

Figure 1 presents the power distribution of the maximally compressed pulse in the temporal domain and the power evolution behaviors with distance at $\tau = 0$ for different modulation amplitudes and widths, respectively. From it one can see that the peak power is increasing as function of both the modulation amplitude $\delta$ and the width $T_1$ [Figs. 1a and 1c], and the peak position is decreasing as function of the modulation amplitude $\delta$ and is increasing as function of the width $T_1$ [Figs. 1b and 1d]. Thus one can control the power and position of the maximally compressed pulse by adjusting suitably the modulation amplitude $\delta$ and the width $T_1$ of the initial Gaussian-type perturbation pulse. Note that for relatively large modulation amplitude $\delta$ and small width $T_1$, the simulations also show the appearance of a secondary localization phase due to the FPU recurrence [35], as shown in Figs. 1b and 1d.

In order to further explain these features, Fig. 2 summarized some results that are useful for us to control the peak power and position of the maximally compressed pulse. Thus Figs. 2a and 2b present the contour lines for the peak powers 8 W, 9 W, and 10 W of the maximally compressed pulse on the $\delta T_1$-plane and the corresponding positions, respectively. For example, if wanting to obtain the maximally compressed pulse with the peak powers 8 W, 9 W, and 10 W at the distance $\xi = 4$ km, we should take the modulation amplitudes and width as $\delta = 0.06$, $T_1 = 6.20$ ps, $\delta = 0.0765$, $T_1 = 8.11$ ps, and $\delta = 0.097$, $T_1 = 9.77$ ps, respectively (see the dashed lines in Figs. 2a and 2b), where the corresponding profiles of the maximally compressed pulses are shown in Fig. 2c.

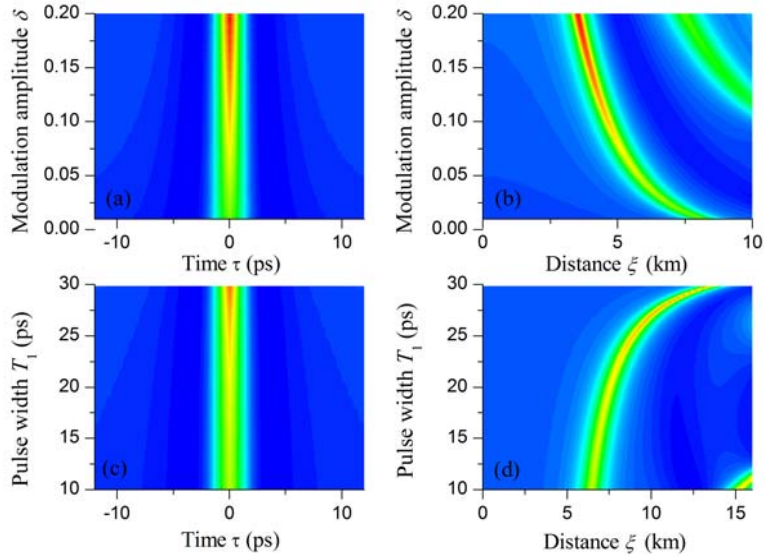

Fig. 1 – Power distribution of the maximally compressed pulse in the temporal domain and power evolutions with distance at $\tau=0$ (a,b) as a function of modulation amplitude $\delta$ for a given width $T_1 = 13.35$ ps, and (c,d) as a function of pulse width $T_1$ for a given $\delta = 0.05$. Here $P_0 = 1$ W and $\varphi = 0$.

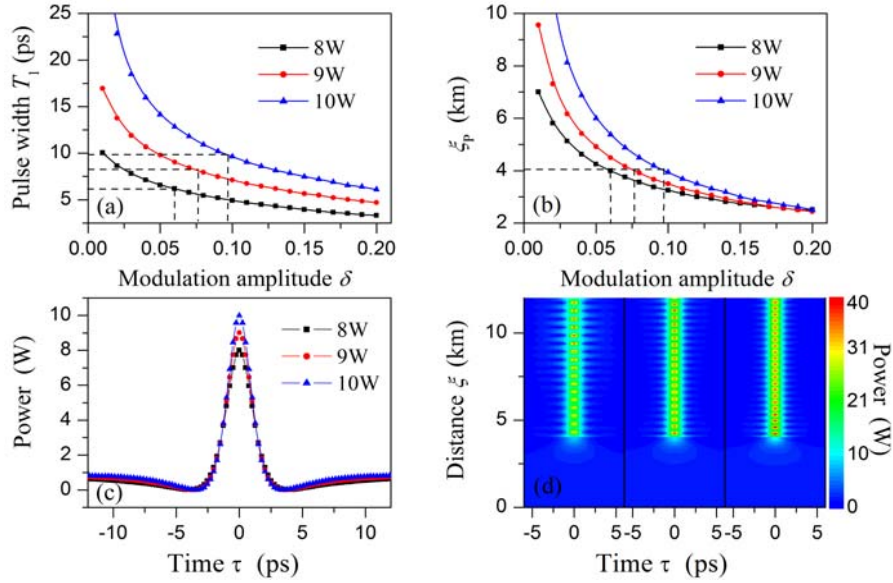

Fig. 2 – a) Contour lines for the peak power of the maximally compressed pulse on $\delta T_1$-plane; b) corresponding peak positions in (a); c) profiles of the maximally compressed pulses with powers 8 W, 9 W and 10 W at the distance $\xi = 4$ km; d) propagation evolution plots of the high power pulses extracted from the maximally compressed pulses shown in (c). Here $P_0 = 1$ W and $\varphi = 0$.

It should be pointed out that the maximally compressed pulse occurred at the peak position can not directly travel with preserving shape, and so can not establish a robust transmission scheme in the optical fiber. This is because of the presence of the background wave at the peak position, which leads to FPU-like growth-return evolution and the pulse splitting due to the modulation instability.

To apply the maximally compressed pulse to generate high power pulses with preserving shape, one must eliminate the background wave $\sqrt{P_0}\, e^{i\xi_P}$ from the maximally compressed pulse. According to this idea, we performed propagation simulations of the evolution of high power pulses by directly eliminating the background wave from the maximally compressed pulses shown in Fig. 2c, and the results are summarized in Fig. 2d. From it one can see that the evolution of the pulses by eliminating the background wave exhibits the oscillatory propagation feature, forming *breathing solitons* with different powers; see Ref. [43] for a previous brief study of this interesting issue.

It should be emphasized that the above mentioned method may be practically implemented by making use of suitable band-stop optical filters. As an example, we consider the maximally compressed pulse with the power 8 W shown in Fig. 2c, and its spectral intensity depicted in Fig. 3a, from which one can see that the spectrum of the background wave concentrate at wavelength 1550 nm. Thus, we can filter the background wave by attenuating the spectra around 1550 nm (about a range of 0.3 nm) to 10% of the corresponding amplitude. Figures 3b and 3c present the corresponding spectral intensity in the spectral domain and intensity profile in temporal domain after filtering the background wave. One can see that the background wave has been successfully eliminated, and we obtain a high power pulse with zero background. Figure 3d presents the propagation evolution plot of the high power pulse extracted from the maximally compressed pulse by employing the spectral filtering method, which is similar to the results shown in Fig. 2d except the oscillatory period becomes larger and the power becomes smaller.

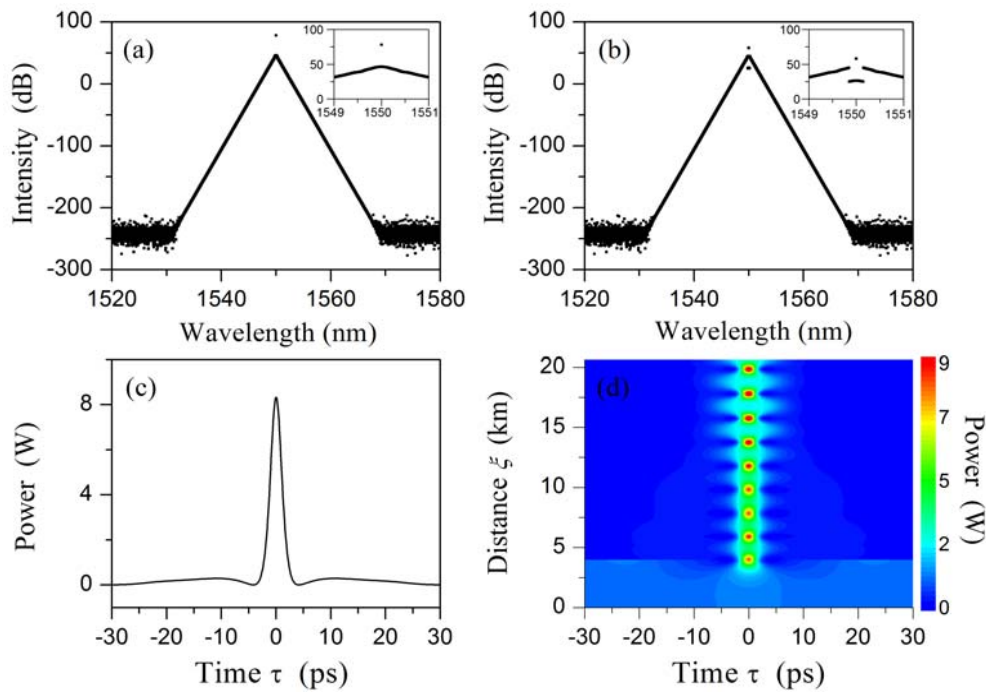

Fig. 3 – a) Spectral intensity for the maximally compressed pulse with power 8 W shown in Fig. 2c; b) the corresponding spectral intensity filtered the spectra around 1550 nm; c) the corresponding profile of the high power pulse with zero background; d) the propagation evolution plot of the high power pulse extracted from the maximally compressed pulse shown in (c). Here $P_0 = 1$ W and $\varphi = 0$.

Next, we discuss the influence of the phase of the initial Gaussian-type perturbation pulse on the dynamics of the maximally compressed pulse. Figures 4a and 4b present the evolution plots of the initial condition (2) for the initial phase $\varphi = 0$ and $\varphi = \pi$, respectively.

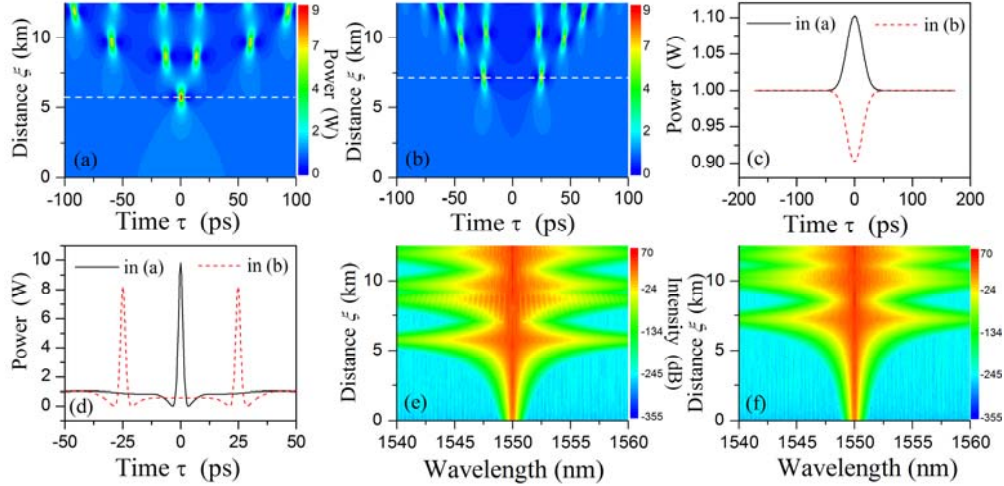

Fig. 4 – a, b) Evolution plots of the initial condition (2) for the different phases $\varphi = 0$ and $\varphi = \pi$, respectively; c) initial profiles; d) intensity profiles at peak position [the white dotted lines in (a) and (b)]; e, f) the corresponding spectral intensities in (a) and (b), respectively. Here $P_0 = 1$ W, $\delta = 0.05$, and $T_1 = 13.35$ ps.

It is surprising that the initial condition (2) with the phase $\varphi = \pi$ can excite a pair of maximally compressed pulses, which differ completely from that with the phase $\varphi = 0$. Also, as a comparison, we present the intensity profiles at the initial position and the peak position, as shown in Figs. 4c and 4d. From them, one can see that their initial profiles are totally different. For the case with the phase $\varphi = 0$, the initial profile is a small localized peak perturbation pulse of CW background, which excites a maximally compressed pulse at the peak position. But when the phase $\varphi = \pi$, the initial profile is a small localized dip on the CW background, which leads to the occurrence of a pair of maximally compressed pulses at the peak position and the profile of each maximally compressed pulse is the same with the case $\varphi = 0$, except for the peak power. Moreover, we also compared the corresponding spectral intensities with the initial phase $\varphi = 0$ and $\varphi = \pi$, as shown in Figs. 4e and 4f. One can see that they have similar spectral evolution, which starts with narrow spectral components and then spreads into a triangular spectrum shape, creating a so-called *supercontinuum*, afterwards shrinks to narrow sidebands with the reduction of spectral components, but the narrow sidebands do not recover to the initial state due to the deviation of the initial condition.

Furthermore, we investigate the intensity profile of the maximally compressed pulse at peak position and the corresponding peak position for the initial phase $\varphi$ ranging from 0 to $2\pi$, as shown in Figs. 5a and 5b. From Fig. 5a, it can be seen that when the initial phase $\varphi$ varies from 0 to $2\pi$, the number of the maximally compressed pulses varies from one to three, and then three pulses switch to two pulses (within the small range of values for $\varphi$), but at $\varphi = 1.583\pi$, they rapidly recover to a single pulse. In this scenario, the separation distance between the two maximally compressed pulses is decreasing with increasing the initial phase $\varphi$. In Fig. 5b, the plot of the peak position shows a cycling behavior. Especially, $\varphi = 0.622\pi$ and $\varphi = 1.583\pi$ are two inflection points, see Fig. 5b.

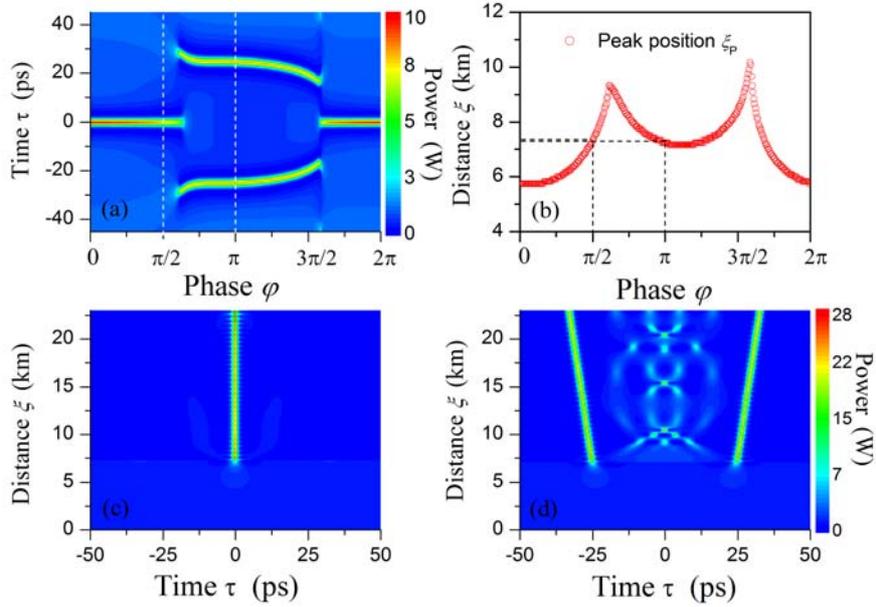

Fig. 5 – a) Distributions of the intensity profile at peak position for different values of initial phase $\varphi$; b) dependence of the peak position on the initial phase $\varphi$; c, d) evolution plots of the high power pulses extracted from the peak position for the initial condition (2) with $\varphi = \pi/2$ and $\varphi = \pi$, respectively. Here the other parameters are the same as in Fig. 3.

Thus we may control the number of the maximally compressed pulses by choosing suitably the phase of the initial Gaussian-type perturbation pulse. As examples, we take the initial condition (2) with $\varphi = \pi/2$ and $\varphi = \pi$ to excite the maximally compressed pulse, where the peak powers of the maximally compressed pulse are 8.20 W at the peak position $\xi_P = 7.32$ km, and 8.26 W at $\xi_P = 7.24$ km, respectively, which are marked with dashed lines in Figs. 5(a) and 5(b). Similarly, we also present the evolution plots of the high power pulses extracted from the peak position, which is obtained by eliminating the background wave $\sqrt{P_0}e^{i\xi_P}$ at the peak position, as illustrated in Figs. 5c and 5d. From Fig. 5d, one can see that the pair of high power pulses exhibits a strong repulsive interaction.

Based on above results, we can obtain robust transmission behaviors of one, two, and even three high power pulses extracted from the maximally compressed pulse by initially exciting a small localized perturbation pulse of the CW background. To obtain an increasing number of high power pulses, we can employ the pulse-splitting effect of the maximally compressed pulse induced by *higher-order modulation instability* [44]. Last, we consider the influence of the width $T_1$ and the modulation amplitude $\delta$ of the initial Gaussian-type perturbation pulse on the pulse-splitting effect of the maximally compressed pulse, respectively. A lot of numerical simulations show that for given initial parameters, the maximally compressed pulse excited by the initial condition (2) can be in turn split into two sub-pluses, three sub-pulses and so on, until becoming irregular.

Figures 6a and 6b present two typical examples, which demonstrate the evolution behaviors of the maximally compressed pulses excited by the initial condition (2) with the parameters $\delta = 0.2$, $T_1 = 22.902$ ps, $\varphi = 0$ and $\delta = 0.16$, $T_1 = 21.115$ ps, $\varphi = 0$, respectively. From them, one can see that the maximally compressed pulse can be at most split into five sub-pulses; that integer number is called the splitting order number, i. e., $N = 5$ in this case.

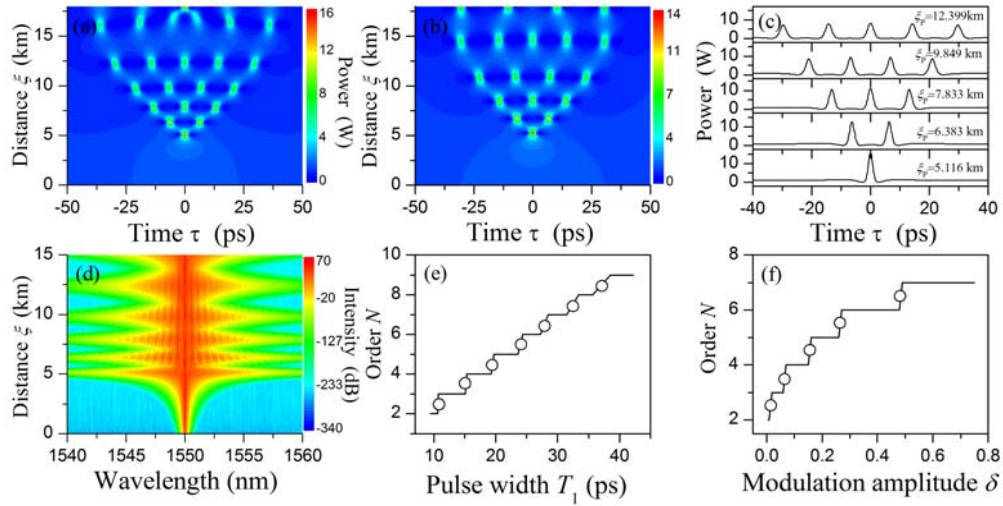

Fig. 6 – Pulse splitting with higher-order modulation instability. For the splitting order number $N = 5$; a,b) the numerical evolution plot of the initial condition (2) with the parameters $\delta = 0.2$, $T_1 = 22.902$ ps, and $\delta = 0.16$, $T_1 = 21.115$ ps, respectively; c) the corresponding intensity profile at different peak positions in (a); (d) evolution of the corresponding spectral intensity for (a); e) dependence of the splitting order number $N$ on the initial width $T_1$ with $\delta = 0.2$; f) on the modulation amplitude $\delta$ with $T_1 = 21.115$ ps. Here $P_0 = 1$ W and $\varphi = 0$.

For the case of Fig. 5a, the corresponding intensity profiles at the different peak positions $\xi_P$ and the spectral evolutions are also shown in Figs. 6c and 6d. Thus, we found that each nonlinear spreading in spectrum is associated with a corresponding pulse breakup. Furthermore, we present the dependence of the splitting order number $N$ on the initial parameters. Figure 6e presents the dependence of the splitting order number $N$ on the initial width $T_1$ for given $\delta$ and $\varphi$, and Fig. 6f presents the dependence of the splitting order number $N$ on the initial modulation amplitude $\delta$ for given $T_1$ and $\varphi$. From them, one can see that the splitting order number is increasing with increasing $T_1$ and $\varphi$, respectively.

Thus, one can adjust the modulation amplitude or the width of the initial Gaussian-type perturbation pulse to control the splitting order number of the maximally compressed pulse. Based on the pulse-splitting effect, we performed the evolution behavior of high power pulses extracted from the maximally compressed pulse, which can be obtained by eliminating the background wave $\sqrt{P_0}\,e^{i\xi_P}$ at every peak position $\xi_P$. Therefore we can establish the transmission scheme for a set of high power pulses. Figures 7a and 7b present the evolution plots of three and four high power pulses extracted from the peak position $\xi_P = 7.833$ km and $\xi_P = 9.849$ km for $N = 5$, respectively. One can see that the high power pulses also exhibit strong repulsive interaction. It should be pointed out that the middle two high power pulses in Fig. 7b exhibit strong interaction. This is because the peak powers of the four pulses split by the maximally compressed pulse are different, as shown in Fig. 7c (see also Fig. 6c), which produce four high power pulses with different peak powers on zero background after eliminating the background wave $\sqrt{P_0}\,e^{i\xi_P}$ at the position $\xi_P = 9.849$ km, as shown in Fig. 7d.

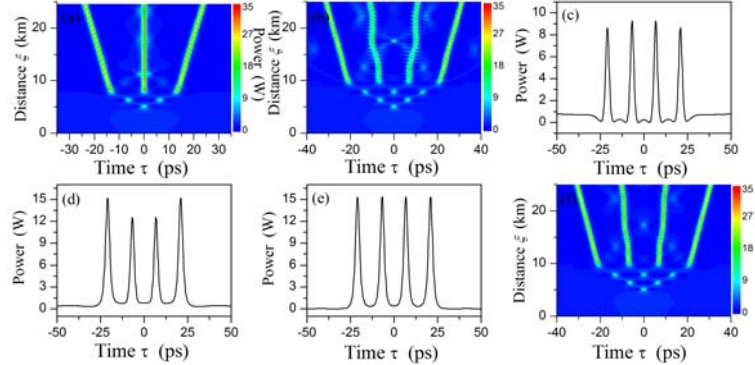

Fig. 7 – a, b) Evolution plots of three and four high power pulses for $N = 5$. For the case of four high power pulses; c) intensity profile before eliminating the background wave; d) intensity profile after eliminating the background wave $\sqrt{P_0}\,e^{i\xi_P}$ with $\xi_P = 9.849$; e) intensity profile after eliminating the background wave $\sqrt{P_0}\,e^{i\xi'}$ with $\xi' = 9.40$, and (f) evolution plot of the four high power pulses by eliminating the background wave $\sqrt{P_0}\,e^{i\xi'}$ with $\xi' = 9.40$. Here $P_0 = 1$ W, $\delta = 0.2$, $T_1 = 22.902$ ps, and $\varphi = 0$.

In order to optimize the result, we performed a lot of numerical simulations and found that by suitably adjusting the phase of the eliminated background wave, we can improve the peak powers of the four high power pulses, as shown in Fig. 7e, which presents the intensity profile after eliminating the background wave $\sqrt{P_0}e^{i\xi'}$, where $\xi'' = 9.40$. From it, one can see that the peak powers of the four high power pulses are now equal. In this scenario, comparing with the result shown in Fig. 7b, the evolution behavior of the four high power pulses has been largely improved, as can be seen from Fig. 7f.

### 3. CONCLUSIONS

In summary, based on nonlinear Schrödinger equation, we have studied a quite simple initial excitation of the maximally compressed pulse and the corresponding generation and transmission of high power pulses extracted from such maximally compressed pulses in a nonlinear optical fiber. The extensive numerical simulations have shown that one can adequately control the power, position, and splitting order number of the excited maximally compressed pulse by suitably choosing the initial perturbation parameters.

Thus one can establish a robust transmission regime of high power pulses extracted from maximally compressed pulse in a fiber communication system. The application of these results in optical fiber systems might be an interesting task at present and in the near future.

*Acknowledgments.* This research was supported by the National Natural Science Foundation of China under Grant No. 61078079 and the Shanxi Scholarship Council of China under Grant No. 2011-010. The work of D. M. was supported in part by Romanian Ministry of Education and Research, CNCS-UEFISCDI, project number PN-II-ID-PCE-2011-3-0083.